\documentclass[%
 aip,
 sd,%
 amsmath,amssymb,
preprint,%
]{revtex4-1}

\pdfoutput=1

\usepackage[english]{babel}

\usepackage{graphicx}
\usepackage{dcolumn}
\usepackage{bm}
\usepackage{hyperref,varioref}
\hypersetup{colorlinks,citecolor=blue,linkcolor=red,urlcolor=blue}
\begin{document}

\preprint{AIP/123-QED}

\title[]{Lipid-protein interaction induced domains:  kinetics and conformational changes in multicomponent vesicles}
\author{K. K. Sreeja}
\email{sreeja@seas.upenn.edu}
\affiliation{Department of Physics, Indian Institute of Technology Madras, Chennai 600036, India, Department of Chemical and Bio-molecular Engineering, University of Pennsylvania, Philadelphia, PA 19104, USA.}

\author{P. B. Sunil Kumar}
\email{sunil@iitm.ac.in}
\affiliation{Department of Physics, Indian Institute of Technology Madras, Chennai 600036, India, Department of Physics, Indian Institute of Technology Palakkad, Palakkad, 678557, India.}

\date{\today}

\begin{abstract}
The spatio-temporal organization of proteins  and the associated morphological changes in membranes  are of importance in  cell signaling.  Several mechanisms that promote  the aggregation of proteins at low cell surface concentrations have been investigated in the past.  We show, using Monte Carlo  simulations,  that the affinity of proteins for specific lipids can  hasten its aggregation kinetics. The lipid membrane is modeled as a dynamically triangulated surface with  the proteins  defined as   in-plane fields  at the vertices.   We show that, even at low protein concentrations, strong lipid-protein interactions can result in large protein clusters  indicating a route to lipid mediated signal amplification.  At high protein concentrations the  domains form buds  similar to that seen in lipid-lipid interaction induced phase separation.  Protein interaction induced domain budding is suppressed  when proteins  act as  anisotropic inclusions and exhibit nematic  orientational order. The kinetics of protein clustering and resulting conformational changes  are shown to be significantly different for the isotropic and anisotropic curvature inducing proteins.\\

\end{abstract}

\pacs{}
\keywords{membrane remodeling, domain formation, lipid-protein interaction, monte carlo simulations}
\maketitle

\section{\label{sec:level1}INTRODUCTION}

 Protein redistribution and clustering  on the cell surface  are important for  signal transduction  pathways~\cite{Simons:2000}.   At its low physiologically relevant cell surface concentrations,  direct interaction between proteins cannot be the primary cause for clustering,  and many active and passive mechanisms, that  indirectly aid protein clustering  have been proposed~\cite{Sarasij:2007, Goswami:2008, Kripa:2012, Raghupathy:2015}.   The specialized membrane domains known as rafts, which is the result of  a sterol and sphingolipid enriched aggregation, are  believed to be one of the precursors for the protein clustering process~\cite{Simons:1997,Simons:2000,Simons:2004}.  Such membrane domains  are often associated  with  peripheral and integral proteins~\cite{Mercker:2015,Barlowe:1994,Antonny:2006,Poveda:2008,Epand:2008,Wu:2014}.   Rafts are known to make a suitable platform for aggregation of GPI(glycosylphosphatidylinositol)- anchored proteins, which correspond to a set of exoplasmic, eukaryotic proteins exhibiting specific intracellular sorting and signaling properties~\cite{Raghupathy:2015,Mayor:2004}.   Caveolin and  clathrin are some of the non raft  proteins associated with lipid domains~\cite{Anderson:2002,Sarasij:2007,Ford:2002}.  Caveolae are  glycolipid enriched domains, that are flask like invaginations  formed by the assembly of Caveolin proteins. It is not clear  if these lipid-protein domains arise from the direct  interactions between the proteins or due to the  interaction between  non-protein membrane constituents  and the  affinity of proteins  to certain membrane composition~\cite{Drucker:2013,Oliva:2015}.    

Another factor  that  has   lead to  considerable interest  in understanding the mechanisms behind lipid-protein sorting in biological membranes is the asymmetric distribution of lipids and proteins in the intercellular organelles such as the golgi and endoplasmic reticulum~\cite{vanMeer:2008, vanMeer:2011}.  Lipids can dynamically vary the constituents of a membrane   by selectively  recruiting various proteins which in turn can change the  functionality of the membrane, and similarly  proteins can sort lipids to specific membrane locations through steric or electrostatic interactions~\cite{McLaughlin:2005,Frost:2009,Bogaart:2011}. 

The membrane protein aggregation due to lipid-protein  interactions has been studied using coarse grained molecular dynamics approaches when the length scale of interest are of few tens of nanometers~\cite{Bogaart:2011}. Since our aim here is to explore  the role of  lipid-protein interaction in the formation of domains and since  the length scale of the resulting conformational changes  are much larger than the membrane thickness, we consider a mesoscale computational approach. Existing computational studies on the     equilibrium  or  dynamic properties of membranes  membranes mostly deals with  lipid  phase separation  following a quench into the two phase coexistence regime~\cite{Kumar:1998,sunil:2001,Laradji:2004,laradji:05}.   Experimental  validation of the results from mesoscale simulations~\cite{Li:2005} has  motivated further  studies on the dynamics of these  domains~\cite{Sanoop:2009}.  However, there are very few attempts to understand how the interaction  of proteins with other membrane constituents  can lead to  compositional inhomogeneities and clustering of proteins.  Here we use a Monte Carlo model for a three component  fluid vesicle, with one component representing protein inclusions and the others two different compositions  of lipids,    to investigate  how lipid-protein interactions affect the kinetics of  domain formation  and associated  conformational changes in the vesicle. To study protein clustering in detail, in our model,  we also account for the direct protein-protein interactions and  the membrane curvature inducing  properties of the proteins. 

  The paper is organized as follows. In Section~\ref{model}, we describe the Monte Carlo model  for a multi-component membrane. In the results and discussions given in  Section~\ref{results},  we first focus on the lipid domain formation. A comparison of the kinetics of lipid-protein interaction induced  domain formation   with that due to lipid-lipid interaction is presented in  Section~\ref{results1}.   Conformational changes  of the membrane for different protein concentrations are also discussed here. In Section~\ref{results11}, we present results on  the effect of  combined  lipid-lipid and  lipid-protein interaction on domain growth. In the second part of Section~\ref{results}, we present our results on  protein  cluster  growth with varying  lipid and protein concentrations.  Section~\ref{results22} and ~\ref{results23} is dedicated, respectively,  to discussion on  how explicit protein-protein interactions and the curvature remodeling  activity of proteins affect cluster growth.

 Our main results are the following.  We show that   the affinity of proteins for certain type of lipids can lead to formation of large protein clusters even at low concentration of proteins. The increase in  protein  cluster growth rate, due to  strong lipid-protein interaction,  indicate a route to lipid mediated signal amplification.   It is pointed out that the absence of line tension at the  domain boundary,  at low   protein  concentration,   is the primary reason for enhanced kinetics of domain growth.  The domain growth is slower at  high protein concentrations and  when proteins  act as  anisotropic inclusions to  exhibit nematic  orientational order. 

\section{Model}\label{model}

         In the Monte Carlo simulations carried out here, the conformation of a  lipid membrane is  approximated to be that of  an elastic sheet represented by  a randomly  triangulated surface.  In this scheme,   a vesicle of spherical topology is represented by $N_v$ vertices, $N_L=3(N_v-2)$ links and $N_T=2(N_v-2)$ triangles and the triangulation is changed randomly to simulate the in-plane fluidity of the membrane.   In the case of an isotropic and homogeneous membrane, the energetics of the elastic sheet is   described using a discretized form of the  Helfrich Hamiltonian,
 \begin{equation}
\mathbf{H}_{\mathrm{elastic}}=\frac{\kappa}{2}{\sum_{v=1}^{N_v}(2H_v-C_0)^2 }A_v,
\label{Helfrich1}
\end{equation}
where summation is over all the vertices of the triangulated membrane. Here $\kappa$ is  the bending rigidity of the membrane,  $C_0$ is the spontaneous curvature resulting from lipid asymmetry of protein induced deformations in the bilayer,  $H_v$ is the mean curvature defined at the vertex  and  $A_v$ is the area  associated with the vertex $v$, computed as in Ramakrishnan et. al.~\cite{Ramakrishnan:10}.  

To model a multicomponent vesicle with two coexisting  lipid compositions and  one type of protein, we introduce a lipid composition field $\phi_v$ and a protein field $p_v=1$, both positioned on the membrane vertices.  The lipid field $\phi_v=1$  when the composition of lipid at vertex $v$  is labelled  A and $\phi_v=-1$ if the lipid composition is labelled  B.   Similarly $p_v=1$ in the presence of a protein at  the vertex and $p_v=0$ otherwise.  A vertex can be simultaneously occupied both by the protein and the lipid fields.  Lipid-lipid interactions are assumed to be Ising like.  The two component lipid vesicle model has been previously used to study the lipid induced  phase separations and budding of domains~\cite{sunil:2001}. The  Hamiltonian describing explicit lipid-lipid  interactions is  given as 
 \begin{equation}  
\mathbf{H}_{\phi}=-J_{\phi} \sum_{\langle vv^{\prime}\rangle}\phi_v\phi_{v^{\prime}}, 
\end{equation}
where the summation runs over all the neighboring pairs. 

 The lipid-protein interaction is modeled using the  Hamiltonian, 
 \begin{equation}
{\mathbf{H}_{p\phi}}=-J_{p\phi}\sum_{\langle v v^{\prime}\rangle' } p_v \phi_{v^{\prime}}.
 \end{equation}
The prime on the summation indicates that  the  protein field at any vertex is allowed to interact with the lipids within the one ring neighborhood including its own vertex. When we choose  $J_{p\phi}>0$,  the  lipid-protein interaction is  attractive  between the proteins and  type A lipids.  Since A lipids are miscible in B lipids and is the minority component,  in this article we will often refer to  lipid composition of type A as  co-lipid. 
 
 \begin{figure}[h!]
\centering
\includegraphics[scale=0.35]{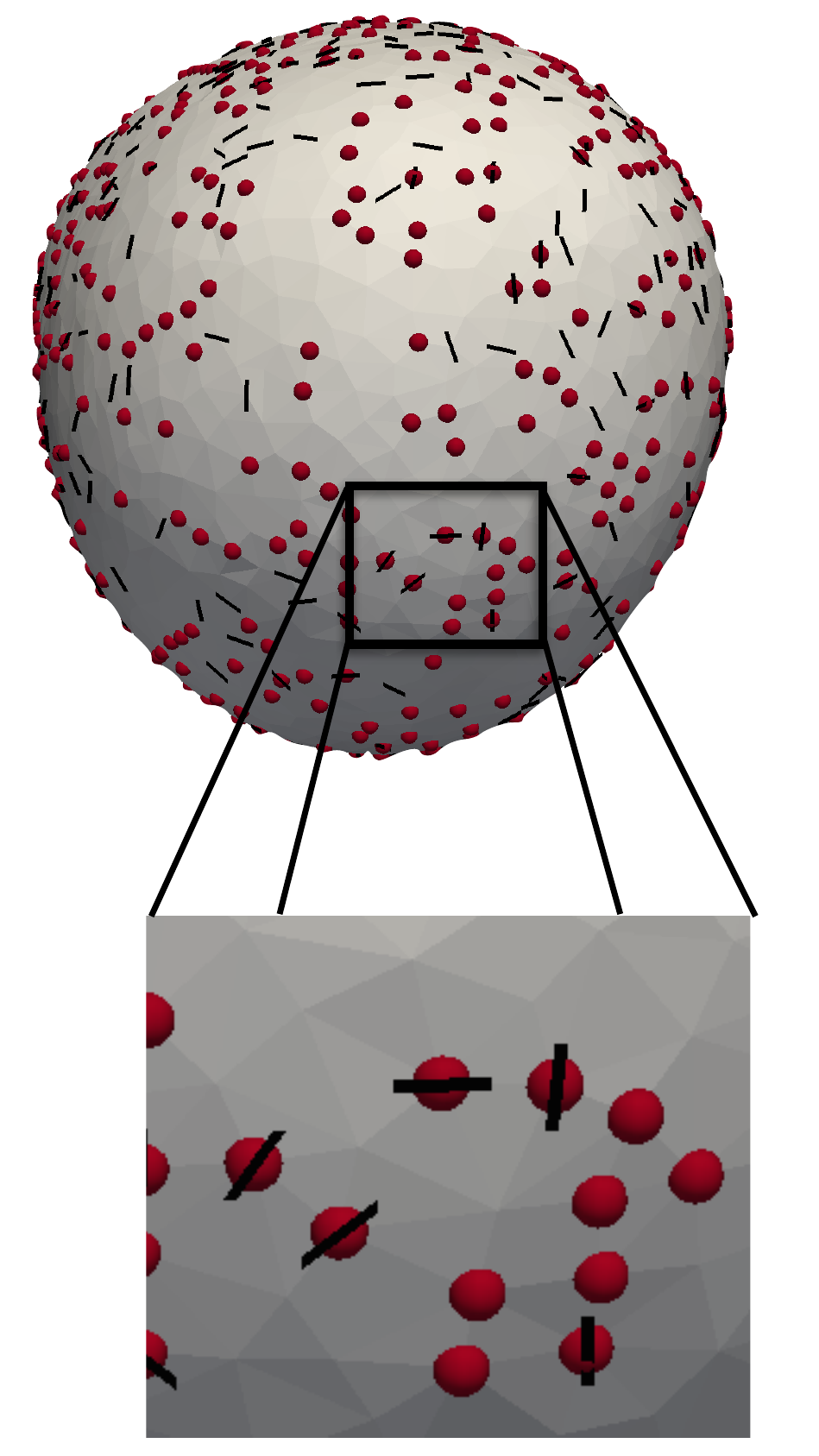}
\caption{ An illustration of a multicomponent vesicle: type A  lipids are  marked as spheres while the unmarked  vertices correspond to type B lipids.  The protein inclusions are shown by black lines.}
\label{nematic_sample}
\end{figure}

   The presence of  curvature active  proteins  modulate local membrane shapes by inducing spontaneous curvature. We study two classes of curvature generating proteins: the first kind of proteins  induce  a  uniform mean curvature  on the membrane and the second kind  are structurally anisotropic proteins, inducing directional curvatures.   To model an isotropic curvature generating protein we consider the spontaneous curvature $C_0$ (see Eqn.\ref{Helfrich1}) at any vertex to be nonzero in presence of a protein. The second class of proteins or protein complexes  have  an extended  structure and cannot be considered as point like objects~\cite{Peter2004}.  To incorporate the structural  anisotropy of such a protein into the model we  introduce a unit orientation vector ${\hat {\bf n}}_v$ such that  the in-plane protein field is now a vector ${\bf p}_v=p_v {\hat {\bf n}}_v$.  The anisotropic shape of the protein  is reflected by the rotational asymmetry of ${\hat{\bf n}_v}$.   In this paper we consider only the case wherein ${\hat {\bf n}_v}$  has a $\pi$ rotational symmetry representing elongated  protein inclusions~\cite{Ramakrishnan:10}.   The  protein field thus acts like a nematic orientational field on the membrane.  The explicit orientational interaction between the  proteins on the membrane is modeled  using the Lebwohl-Lasher model~\cite{Lebwohl:1972}, given by

  \begin{equation}
\mathbf{H}_{LL}=- \epsilon_{LL}\sum_{\langle vv^{\prime}  \rangle} p_vp_{v^{\prime}}\Big\{\frac{3}{2}\cos^2\Phi(\hat {\bf n}_v, \hat {\bf n}_v^{\prime})-\frac{1}{2}\Big\},
\label{LL}
\end{equation}
where, $\Phi(\hat {\bf n}_v,\hat {\bf n}_v^{\prime})$,  the angle between the two in-plane field  vectors on the tangent planes at neighboring vertices $v$ and $v^{\prime}$,  is computed using a parallel transport operation~\cite{Ramakrishnan:10}.   The protein orientation field   is coupled to the  membrane curvature using a discretized version of the Hamiltonian proposed by Frank and Kardar~\cite{Frank:2008},
 \begin{equation}
\mathbf{H}_{\mathrm{anis}}={\sum_v\Big[\frac{\kappa_{\parallel}}{2}(H_{n,\parallel}-C_0^{\parallel})^2+\frac{\kappa_{\bot}}{2}(H_{n,\bot}-C_0^{\bot})^2\Big]}A_v.
\label{hanis} 
\end{equation}
$H_{n,\parallel}$ and $H_{n,\bot}$ are the curvatures along the directions  $\hat{\bf{n}}_v$ and $\hat{\bf{n}}_v^{\bot}$ respectively. $C_{0}^{\parallel}$ and $C_{0}^{\bot}$ are the  local directional spontaneous curvatures and $\kappa_{\parallel}$ and $\kappa_{\bot}$ are the bending stiffness along $\hat{\bf{n}}_v$  and $\hat{\bf{n}}_v^{\bot}$,  respectively. The anisotropic protein inclusions considered in our study generate  additional stiffness and curvatures only  along the direction $\hat{\bf{n}}$, i.e., we consider only the cases with  $\kappa_{\bot}=0$.

   The multicomponent vesicle is equilibrated through a set of Monte Carlo (MC) moves with the total effective Hamiltonian:
 \begin{equation}
  {\mathbf{H}_{\mathrm{total}}}={\mathbf{H}_{\mathrm{elastic}}}+{\mathbf{H}_{p\phi}}+{\mathbf{H}_{\phi}}+{\mathbf{H}_{\mathrm{anis}}}+{\mathbf{H}_{LL}}.
 \end{equation}
 We only consider the case of conserved in-plane fields and the  MC moves  include the vertex moves, link flips, {\bf p} field rotations  and the  exchange of  ${\bf p}$ and $\phi$ fields ~\cite{Ramakrishnan:10, Ramakrishnan:2013}. Unless otherwise stated, all moves are accepted through the Metropolis algorithm.  (a) {\it{Vertex move:}} Here the position of a vertex is updated to a new position  chosen randomly  within  a cutoff distance. This cutoff is fixed  such that   $50\%$ of the moves are accepted. This move allows for shape changes of the vesicle. (b) {\it{Link flip:}} A randomly selected edge, connecting two triangles, is disconnected and a new connection between the unconnected vertices of the same triangles is constructed. This move changes the triangulation/connectivity and physically models the fluidity of the bilayer membrane by allowing the vertices to diffuse through the surface. (c) {\it{Exchange of $\phi$ fields:}} The diffusion of lipid composition field  on the membrane surface is captured using a Kawasaki move that allows for an exchange between type A and type B vertices. (d) {\it{Exchange of {\bf p} fields:}} The diffusion of protein field on the membrane surface is captured using a Kawasaki exchange move. (e) {\it{{\bf p} field rotation:}} In this step, a vertex is chosen randomly and the orientation  field at the chosen vertex, if nonzero, is rotated to a new, randomly chosen direction in the tangent plane. This rotation of the field allows for the relaxation of the orientational order of the field.

The multicomponent membrane system described here is studied using vesicles with 2030 vertices and 4056 triangles. We consider a vesicle with  bending rigidity  $\kappa=10$ $k_BT$. Initial configurations of the vesicle are generated by randomly assigning  $\phi_{\%}$ of the membrane vertices to have lipids of type A and  the rest of the vertices are assigned to have type B lipids. The  proteins, whose number fraction is represented by  $p_\%$,   are also placed  at  randomly chosen vertices with random in-plane orientations. A patch of the membrane with  co-existing lipid and protein fields is shown in Fig.~\ref{nematic_sample} and we follow the same representation for further discussions. It should be noted that even when the proteins are not anisotropic in nature they are represented by solid black lines in order to distinguish them from  lipid-type specification on the vertices.

\section {Results and Discussions}\label{results}
 
\subsection{  Lipid clustering due to protein-lipid interaction }\label{results1}  

      To investigate  membrane   inhomogeneities induced by the lipid-protein interactions, we first consider the case of isotropic inclusions, such that   ${\mathbf{H}_{\mathrm{anis}}}$ and ${\mathbf{H}_{LL}}=0$, no direct  lipid-lipid interactions (i.e. $\mathbf{H}_{\phi}=0$)   and  focus on the aggregation kinetics that is solely driven by   lipid-protein interaction  $\mathbf{J}_{p\phi}>0$. For these parameters, the proteins  only serve to promote the aggregation of lipids and  do not have any direct impact on local membrane shapes.   In our model the lipids and proteins can occupy the same vertex and lipid-protein interaction is limited to nearest neighbor sites.  The relative values of $\phi_\%$ and $p_\%$  is thus another important parameter.  Below we first analyze the  cluster growth when $p_\% <\phi_\%$. 

\begin{figure}[h!]
\centering
\includegraphics[scale=0.55,angle=0]{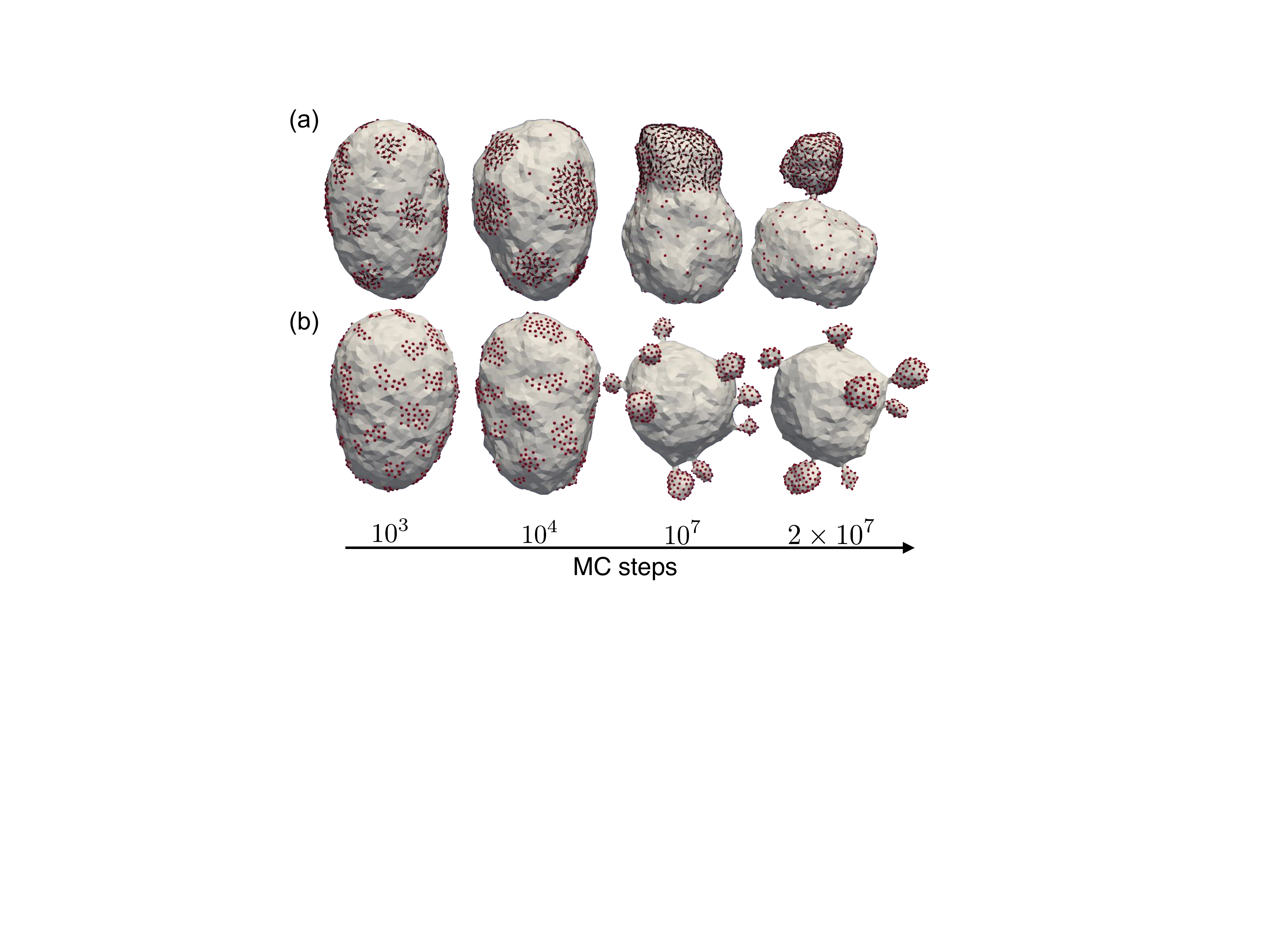}
\caption{A comparison of domain formation with $p- \phi$ and $\phi-\phi$  interactions as a function of MC steps. (a) Conformations of a vesicle when lipids  aggregate through  $p - \phi$ interactions for $J_{\phi}=0$, $J_{p\phi}=2$, $\phi_\%=30$ and $p_\%=20$. (b)  Lipid clusters induced by  $\phi$ - $\phi$ interactions for $J_{\phi}=2$, $\phi_\%=30$ and $p_\%=0$.  }
\label{conf1}
\end{figure}

{\it{Clustering kinetics:}}  The  evolution of  membrane inhomogeneities and the associated vesicle  conformations,  as a function of MC time, are shown in Fig.~\ref{conf1}.    Panel (a) shows $p-\phi$  interaction driven clustering when $\mathbf{H}_{p\phi}>0$, $\mathbf{H}_{\phi}=0$, $\phi_\%=30$ and $p_\%=20$. For comparison  panel (b) shows membrane conformations for lipid-lipid interaction induced aggregation, when  $\mathbf{H}_{p\phi}=0$, $ \mathbf{H}_{\phi}>0$, $\phi_\%=30$ and $p_\%=0$.  As can be inferred  from the figure, the kinetics of clustering  and conformational changes in the membrane are significantly different in these two cases. The  evolution of vesicle shape, resulting from direct lipid-lipid interactions, shown in  panel (b) of  Fig.~\ref{conf1}, is similar to that observed in  previous studies~\cite{sunil:2001}.  At early times  ($\mathrm{MC}$ $\mathrm{steps}\leq10^4$), in both cases,  domains of co-lipids nucleate and  grow into stable clusters as shown in Fig.~\ref{conf1}.  As can be seen in panel (a), $p-\phi$ interaction is sufficient to  induce  an effective attraction between the lipids and  trigger the formation of co-lipid  patches. The late time growth of domains, however, is strongly dependent on the nature of interactions.  When $\phi_\%>p_\%$, in the case of  protein induced segregation,   shown in panel (a), the domains remain flat and their  aggregation is fast.  While in  panel (b), the presence of  explicit lipid-lipid interaction  leads to budding and slowing down of coarsening. 

\begin{figure}[h!]
\centering
\includegraphics[scale=0.95]{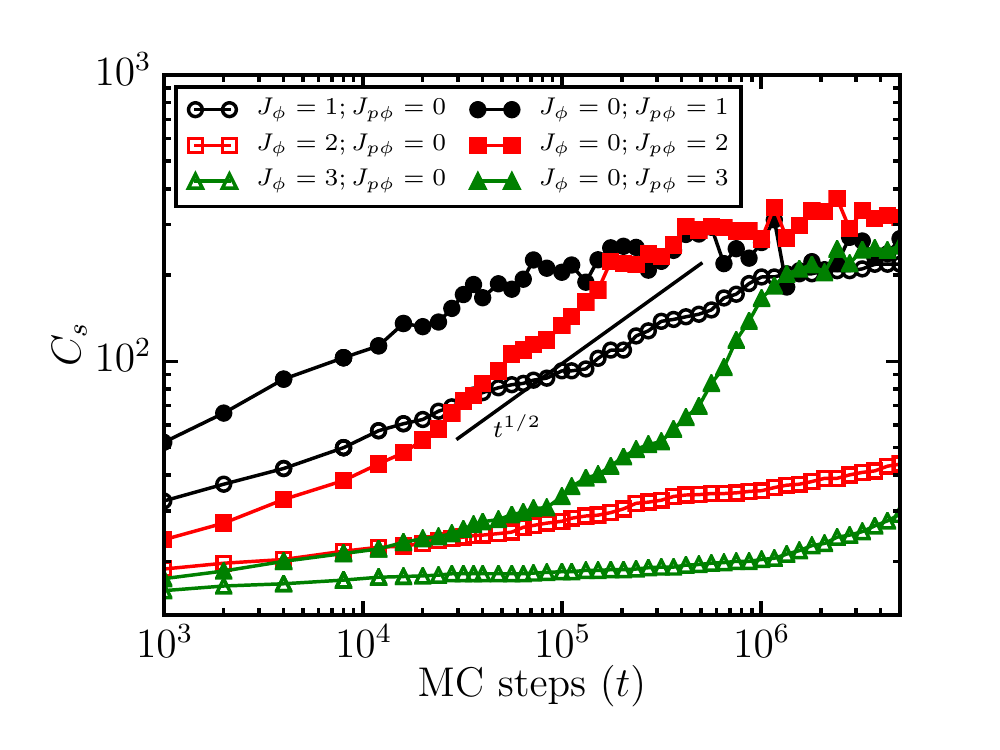}
\caption{Comparison of  lipid cluster growth with $p-{\phi}$  and $\phi-\phi$  interactions when $\phi_{\%}=30$. Average cluster size of type A domains for different values of $J_{\phi}$ and $J_{p\phi}$  is given. When $J_\phi= 0$ (filled symbols) protein concentration is fixed at $p_\% = 20$ while for $J_\phi > 0$ (open symbols), it is taken to be $p_\%$ = 0.}
\label{graph1}
\end{figure}

\begin{figure}[h!]
\centering
\includegraphics[scale=0.28]{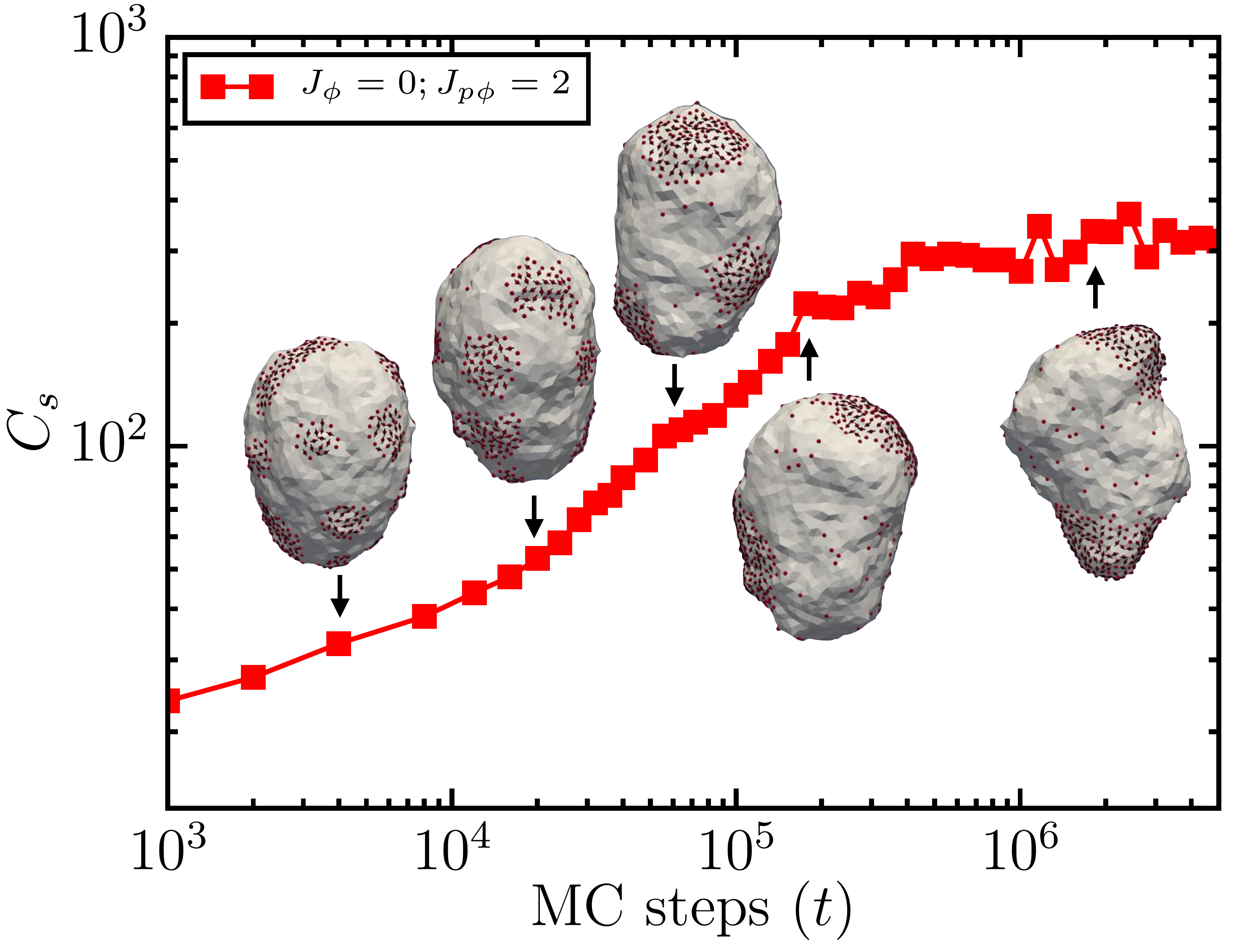}
\caption{Coalescence  of protein rich domains on the vesicle surface due to $p-\phi$ interactions. The cluster sizes shown are for $\phi_{\%}=30$, $p_{\%}=20$, $J_{p\phi}=2$ and $J_{\phi}=0$. }
\label{graph}
\end{figure}

\begin{figure}[h!]
\centering
\includegraphics[scale=0.95]{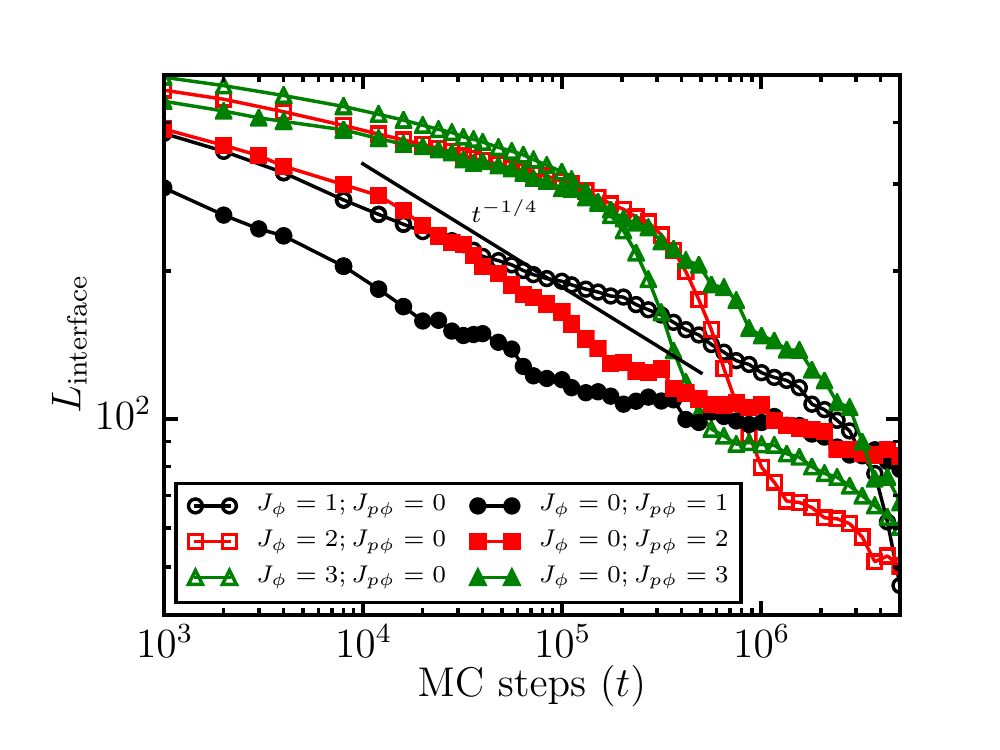}
\caption{Comparison  of lipid domain interfacial length for $\phi-\phi$ and $p-\phi$ interactions for various values of  $J_{\phi}$ and $J_{p\phi}$ when  $\phi_\%=30$.  When $J_\phi= 0$ protein concentration  $p_\% = 20$ and $L_{\mathrm{interface}}$ is computed as the number of vertices occupied by   the A lipids with atleast one type $B$ lipid vertex as a neighbor.}
\label{graph2}
\end{figure}

 A quantitative  comparison of domain growth can be obtained by analyzing the average lipid cluster size  as a function of time. The average cluster size  for different values of the interaction strengths $J_{\phi}$ and $J_{p\phi}$ are given in Fig.~\ref{graph1}.  The cluster sizes ($C_s$)  correspond to the number of vertices of  type A that form a continuous map on the triangulated surface.    The early  time domain growth is similar in all cases.  When the coarsening is only through   lipid-lipid  interaction strength (i.e., for $J_{\phi} >0$ and $J_{p\phi} =0$),  for all values of  $J_{\phi} $,  and for  the system sizes considered here,  the growth rate remains low and the  system does not enter into a scaling regime.  The coarsening is considerably faster  with lipid-protein interaction  $J_{p\phi}>0$  and we see a power law  regime  with  $C_s \propto t^{1/2}$,  which could result from domain diffusion snd coalescence (see~ Appendix \ref{appendix1}), as the diffusion coefficient of domains, in the Rouse dynamics,   varies   as  the inverse of the domain size (data not shown).     The corresponding configurations are shown in  Fig.~\ref{graph}.  It is important to note that,   for $J_{p\phi} \le 3$,  $J_{\phi}=0 $  and with $\phi\%>p\%$, the domains remain flat. This  is evident from the     $Ls \propto t^{1/4}$  dependence of  interfacial length $Ls$  on time  as shown in Fig.~\ref{graph2}.  On the other hand,  in the case of clustering  induced by direct  lipid-lipid  interactions,  when the value of $J_{\phi}$ is higher,  the line tension is usually significant enough to induce budding.   This regime can be easily identified  in  Fig.~\ref{graph2}  as  one  with sudden fast decrease of  interface length.  The movement of domains,  which now involve membrane shape changes,  significantly  slows down domain coarsening.

It is clear from the above discussions that  the ability of domains to remain flat is important  for fast clustering of proteins.   In the model,  positive values of $J_{p\phi}$  favor type A lipids to occupy a vertex with  a protein on it  or in the one ring neighborhood of it.   When the fraction of  vertices occupied by the protein is much smaller than that of A lipids ($\phi_\%>p_\%$),   there are enough   A lipids  to occupy  the one ring neighborhood  even  when  there are many small   protein  clusters.   In this regime, when $J_{\phi}=0$,  co-lipids at the boundary of the domains will act as surfactants and we expect the interfacial tension at the lipid-protein domain boundaries  to be  negligible resulting in flat domains.    Such a surfactant lined domain could   stabilize  small clusters and prevent coarsening.  But our simulations show  a complete  coalescence  of domains,  indicating that the entropy gain from release  of   excess co-lipids  is significant enough to drive domain coarsening. Thus the  protein induced  domains    show faster coalescence  compared to the domains formed by direct  lipid-lipid  interaction.

\begin{figure}[h!]
\centering
\includegraphics[scale=0.45]{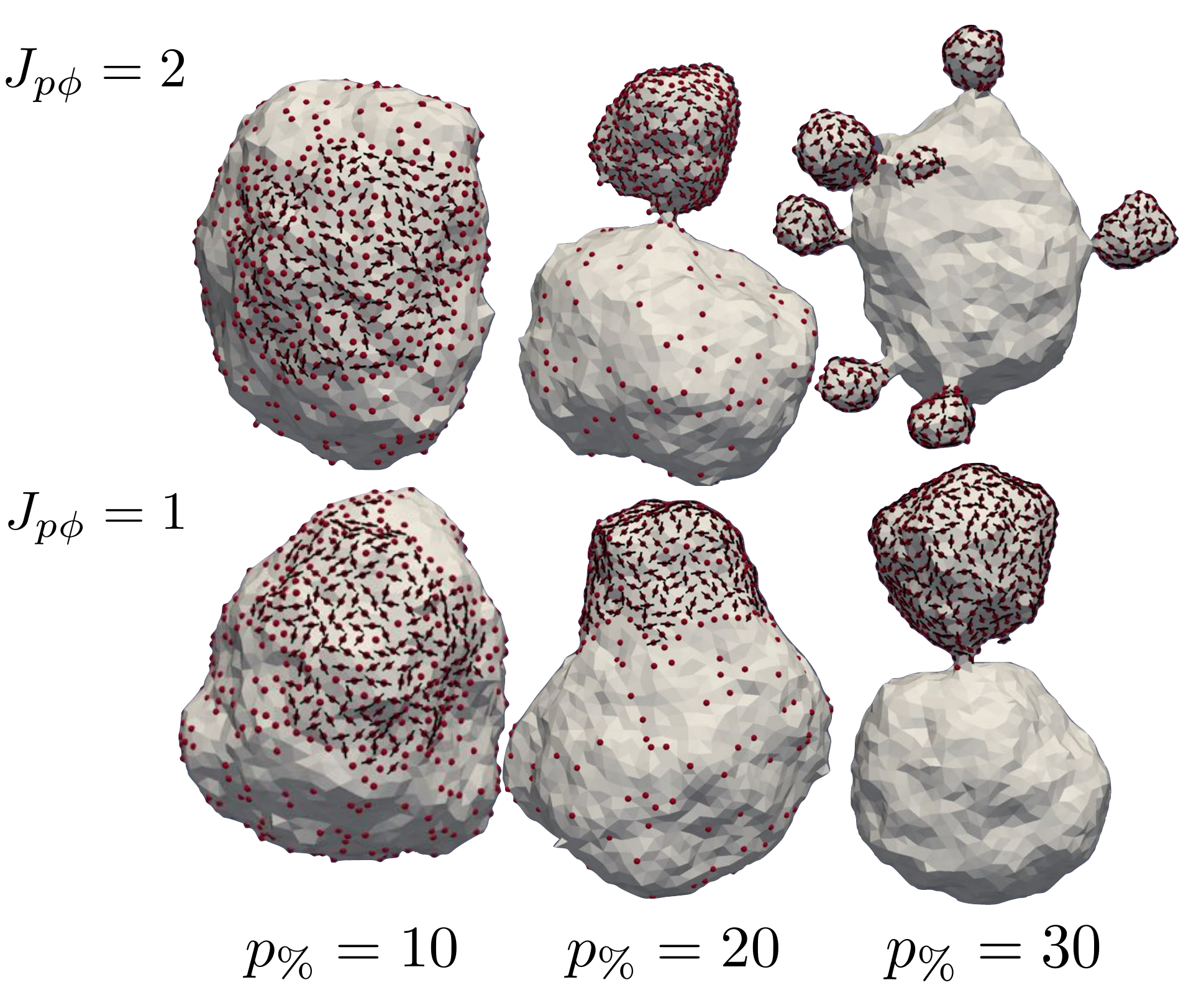}
\caption{Phase separation and shape changes as a function of $p_\%$ and $J_{p\phi}$ when $\phi_\%=30$ and $J_{\phi}=0$. Lipid domains starts to bud when $p_\%=20$ and $J_{p\phi}=2$.  The snapshots are taken  after $2\times10^7$ MC steps.}
\label{conf2}
\end{figure}

 {\it{Conformational changes:} }   When $J_{\phi}=0$,  there are two main  factors that  affect  the conformational changes of the membrane;  (i) the fraction of vertices occupied by the  proteins ($p_\%$)  in comparison to that occupied by  co-lipids ($\phi_\%$) and (ii) the interaction strength $J_{p\phi}$. Fig.~\ref{conf2} shows the representative equilibrium conformations  for different values of  $p_{\%}$   and $J_{p\phi}$ for a fixed value of $\phi_\%=30$.  When  $p_\%=10$, large clusters are formed, but these clusters do not initiate budding even for large value of $J_{p\phi}$.  As described in the previous section, this is due to the presence of  additional type A vertices at the interface that shields the protein from type B vertices and  reduces the line tension.  The domains start to deform  when $p_\%=20$  and  $J_{p\phi}=2$. At $p_\%=30$ the number of proteins become equal to the number of type A vertices,  and there are no additional  co-lipids to line the domain interface. In this case when the lipid-protein domain size reaches a certain value it  starts to bud.        
 
\subsubsection{ Effect of  lipid-protein interaction at low co-lipid concentration}\label{results11}

\begin{figure}[h!]
\centering
\includegraphics[scale=0.9]{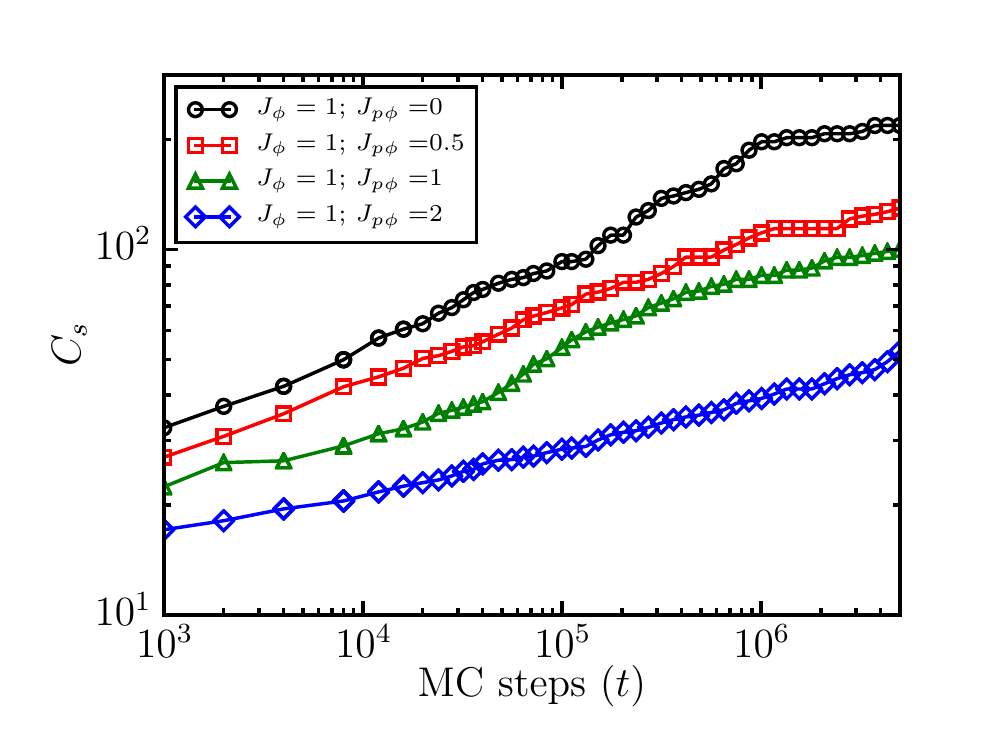}
\caption{Average cluster size with lipid-protein and lipid-lipid  interactions when $\phi_{\%}=30$ and $p_{\%}=20$ for $J_{\phi}=1$ and $J_{p\phi}=0,0.5,1$ and $2.$}
\label{grap3}
\end{figure}

     In biological membranes both lipid-lipid and lipid-protein interactions are expected to affect the protein phase-segregation and hence a study of combined
$\mathbf{H}_{\phi}$ and $\mathbf{H}_{p\phi}$ interactions are of relevance to cellular membranes. Here we study the effect of lipid-protein interactions by holding the  lipid-lipid interaction fixed at $J_{\phi}=1$. The average cluster size in this case when $\phi_{\%}=30$ and $p_{\%}=20$ is shown in Fig.~\ref{grap3}.  Here  as  $J_{p\phi}$ increases the   growth rate of  lipid cluster size decreases in a  monotonic fashion. When  there is not enough co-lipids to cover all protein neighborhood  and the  protein  domain boundary is occupied by both A and B lipids, non zero protein B lipid interaction, parameterized through $J_{p\phi}$, will increase the line tension at the boundary, resulting in domain shape changes. 
 Such morphological changes  reduces domain  diffusion  and  leads to a slow down in coalescence of domains and increases the life time of smaller domains.

\subsection{ Protein clustering due to lipid-protein interaction}\label{results2} 

In this section we investigate how the lipid-protein interaction can change the effective protein clustering and the structural properties of protein inclusions  affect  protein clustering. In the following  discussion we consider domain formation without explicit lipid-lipid interactions and constant lipid-protein interactions, i.e.,  we take $J_{\phi}=0$  and $J_{p\phi}=2$. 

\subsubsection{ Concentration of proteins and lipids}\label{results21} 

  \begin{figure}[h!]
\centering
\includegraphics[scale=0.6]{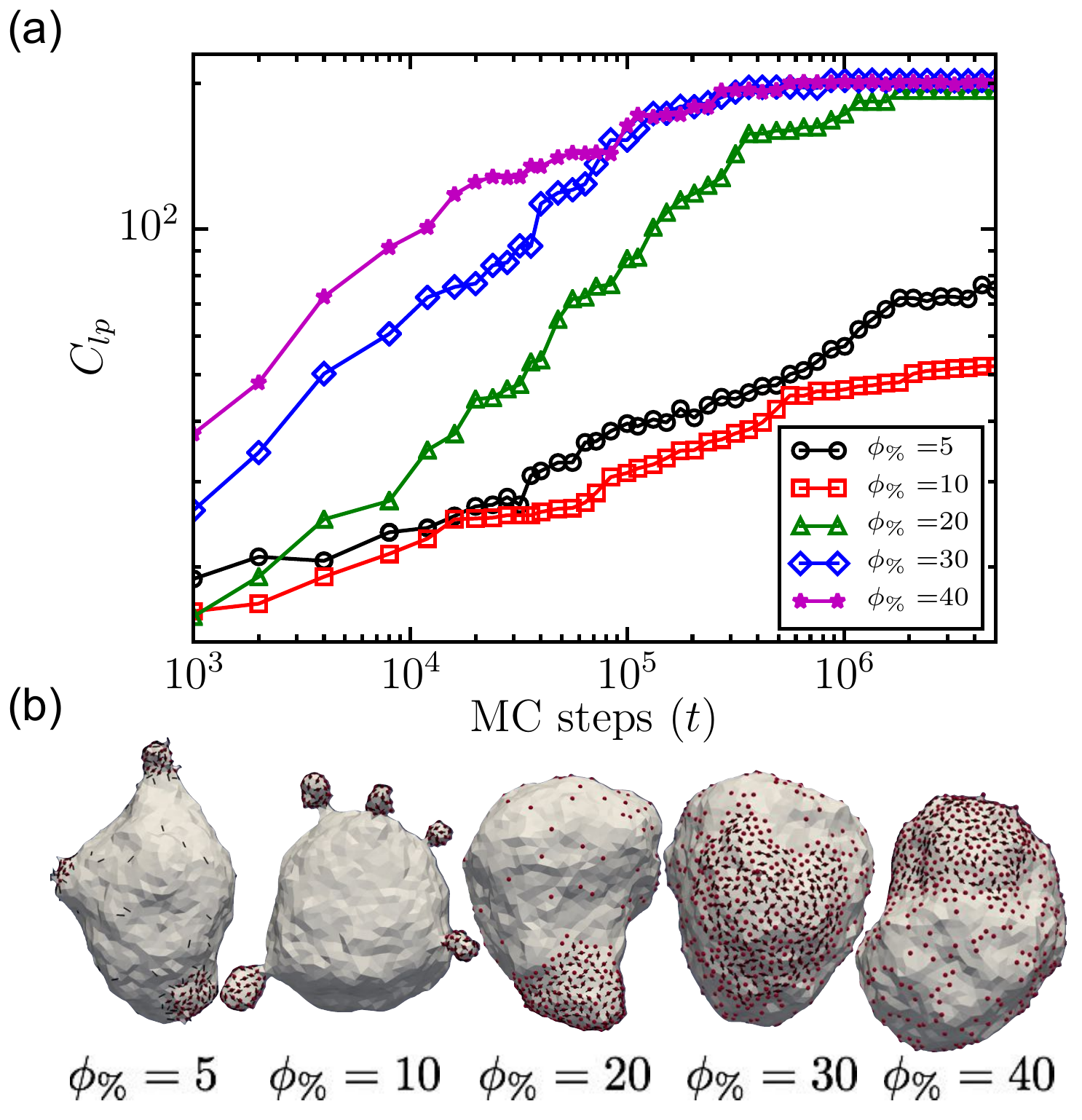}
\caption{ Protein aggregation at  low protein concentration as a function of $\phi_\%$ for $p_\%=10$, $J_{\phi}=0$  and $J_{p\phi}=2$. Panel (a) shows  the size of largest protein cluster as a function of time.  Panel (b) shows  representative vesicle conformations at MC steps$= 5\times 10^6$.}
\label{smallp}
\end{figure}

  Here we study the aggregation of proteins  at a smaller protein concentration compared to  the previous cases. We keep  $p_\%=10$ and  study the domain formation by varying   $\phi _\%$. The resulting conformations and the largest cluster sizes are shown in Fig.~\ref{smallp}.  As shown in panel(b), the domains bud for equal concentration of co-lipids and proteins.  When $\phi_\%> p_\%$ the line tension decreases and hence the domains remain flat. The increase in cluster sizes due to the flat domains can be seen in Fig.~\ref{smallp}(a).  When $\phi_{\%} \ge 20$,  at early times  we observe a domain growth that depends on the fraction of  co-lipids present and  at late times a complete clustering of proteins. Even when  $\phi_\%>>p_\%$  we observe a large cluster instead of small clusters. This could be due to the increase in entropy due to the configurational freedom of proteins in a larger cluster compared to small domains.

 \begin{figure}[h!]
\centering
\includegraphics[scale=0.6]{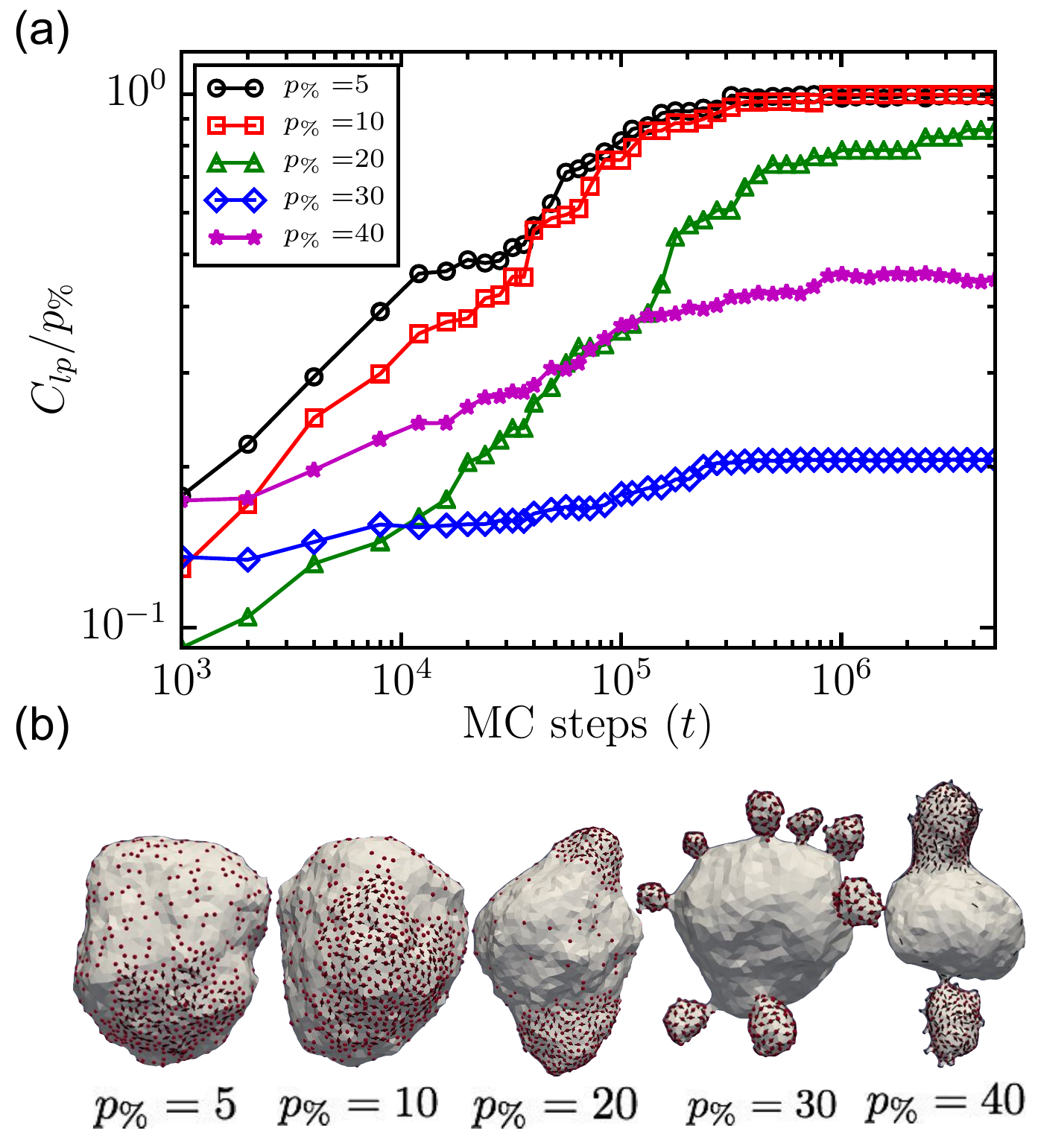}
\caption{ Protein clustering as a function of $p_\%$ when $\phi_\%=30$, $J_{\phi}=0$   and $J_{p\phi}=2$.  Panel (a) shows $C_{lp}/p_\%$ as a function of time and  (b) the vesicle conformations at MC steps$= 5\times 10^6$.}
\label{varyp}
\end{figure}

 Complete protein clustering, even at low values of protein concentration and in the absence of any direct  protein-protein interaction, can be achieved through lipid-protein interaction is one of the main results presented in this paper.   To understand this more quantitatively  we also  looked  at  lipid-protein  aggregation as a function of $p_\%$ when $\phi_{\%}=30$.  In   Fig.~\ref{varyp}(a) we show $C_{lp}/p_\%$ to represent the fraction of proteins clustered and in  Fig.~\ref{varyp}(b),  the corresponding conformations at  $5\times 10^6$ MC steps.  For low protein concentrations ($p_\%=5$, $10$) a complete clustering of proteins is observed. The fraction of protein clustered is minimum when $p_{\%}=30$, which is equal to $\phi_\%$. This is expected as the line tension and membrane deformations are maximum when $p_{\%}=\phi_\%$.
  

\subsubsection{Explicit protein-protein orientational interactions}\label{results22} 

   In this section we focus on the effect of a direct  interaction between the proteins  in addition to  its interaction with lipids. Orientational interaction between proteins are relevant for elongated protein inclusions. The proteins  are now  treated as in-plane vector fields and their   nematic like orientational interaction is  modeled through  the Lebwohl-Lasher energy functional~\cite{Han:2008ck}, which is given in Eqn.~\ref{LL}.  This  interaction is short ranged and is  only between in-plane fields (proteins) within the connected neighborhood and captures the symmetry property of the nematic vectors. In an earlier study it has been shown  that  this orientational interaction ($\mathbf{H}_{LL}$)  alone  can aggregate proteins on the surface and induce significant membrane conformational changes~\cite{Ramakrishnan:10}.

\begin{figure}[h!]
\centering
\includegraphics[scale=0.5]{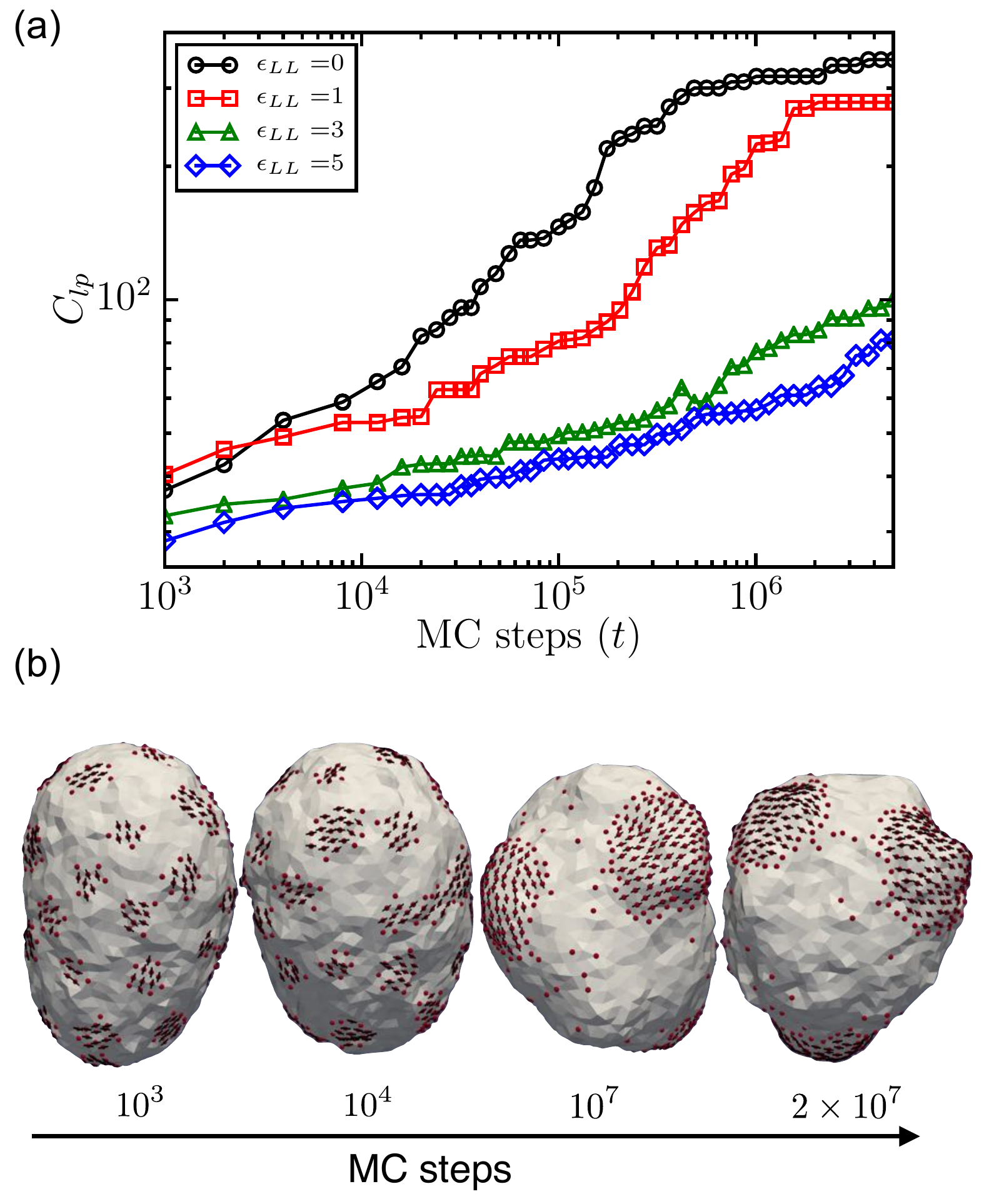}
\caption{Formation of protein clusters with explicit  protein-protein interaction when $J_{p\phi}=2$, $J_{\phi}=0$,  $p_\%=20$ and $\phi_{\%}=30$. Panel (a) is the  largest protein cluster size  and  (b) the vesicle conformations when $\epsilon_{LL}=3$.}
\label{graph3}
\end{figure}

    In  Fig.~\ref{graph3} we  show the time evolution of protein domains on the vesicle with orientational order and fixed $\mathbf{H}_{p\phi}$ interaction; panel (a) shows the largest  protein cluster size and panel (b) the vesicle conformations as a function of time when $\epsilon_{LL}=3$.  We consider the case with $p_\%=20$ and $\phi_{\%}=30$, to compare with the case discussed in Sec~\ref{results1}.   From Fig.~\ref{graph3}(a) it can be seen that  as $\epsilon_{LL}$ increases  the cluster growth slows down. As $\epsilon_{LL}$  increases the  additional attractive interaction between the components reduces the diffusivity of the protein cluster  which in turn reduces the rate of  growth of clusters.  When $\epsilon_{LL}=0$ (see panel (a) of Fig.~\ref{conf1}), $\mathbf{H}_{p\phi}$ interaction induces domains that are nearly circular in shape which, at a later stage, coalesce and bud off.   When $\epsilon_{LL}>0$ (see panel (b) of Fig.~\ref{graph3}), domains with non-circular boundaries are formed first.   Within a domain the protein fields  are aligned in one direction.  As time progress  the  domain boundaries take a  circular shape. However,   due to the  energy cost in  maintaining  parallel orientation of the protein field, the domains  do not  bud as  in  the case with   $\epsilon_{LL}=0$.

\subsubsection{Curvature active proteins }\label{results23} 
   Collective interactions of curvature generating membrane associated proteins, like BAR domain proteins, caveolin and clathrin   can lead to interesting shape transformations in organelle membranes.   Protein induced  lipid sorting has been shown with compelling evidences in the  trans-golgi and endosomal membranes and it is believed that sorting happens in response to the curvature induced by the proteins~\cite{Anderson:2002,Sarasij:2007,Ford:2002,Ungewickell:2007}. The presence of  curvature inducing and curvature sensing proteins in lipid domains can significantly  alter  the composition of the domain since these proteins are known to have a strong affinity for certain class of lipids~\cite{Epand:2005,Mattila:2007, Zhao:2013}.  Here  we discuss how the protein field induced curvature along with lipid-protein interactions, can alter the protein domain growth and the shape of the vesicle. The domain formation due to two different classes of proteins: isotropic and anisotropic curvature generating proteins, are considered.

\begin{figure}[h!]
\centering
\includegraphics[scale=0.5]{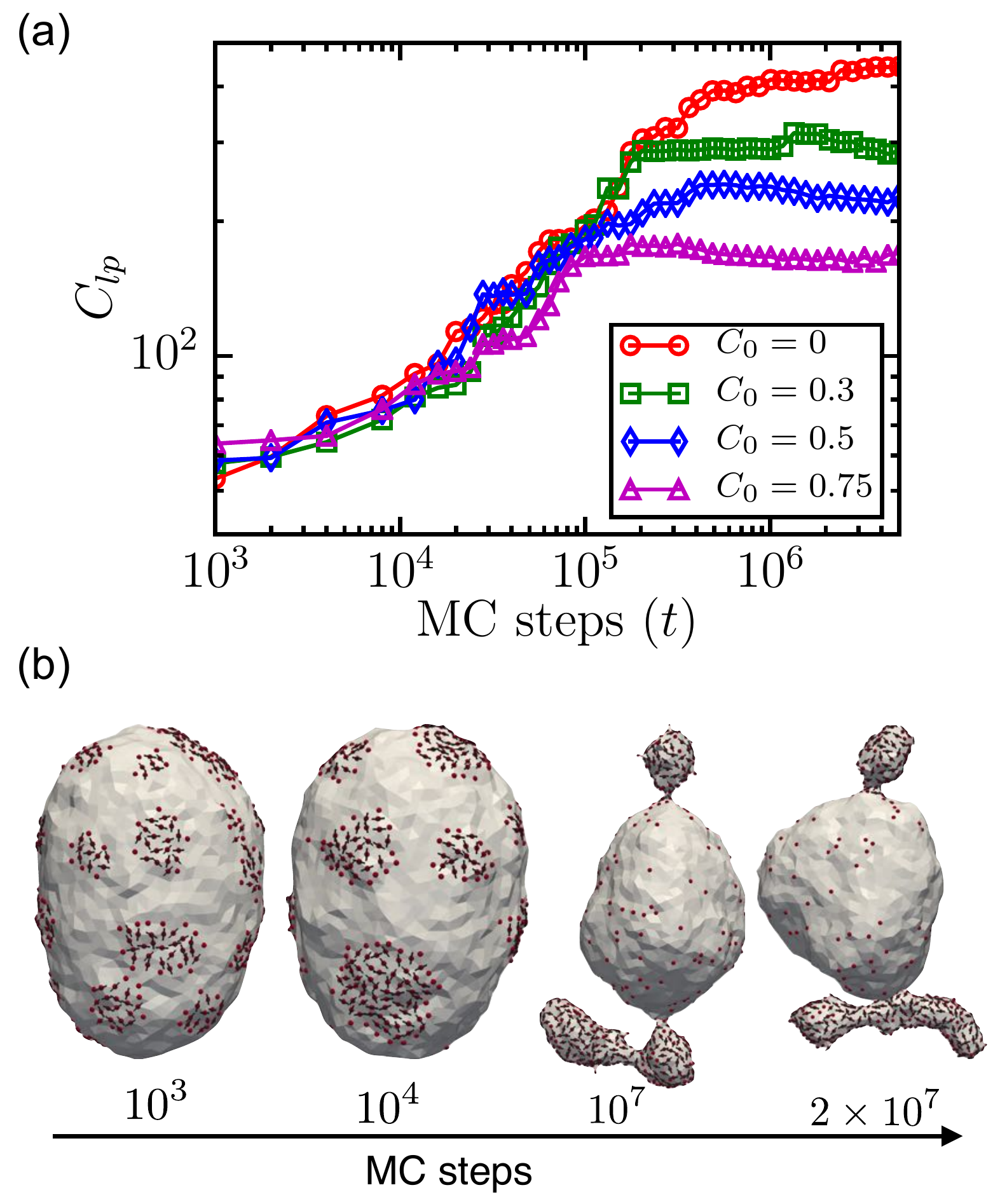}
\caption{Cluster formation in the presence of isotropic curvature generating proteins as a function of time when  $J_\phi=0$ and $J_{p\phi} =2$. Panel (a) shows the largest protein cluster size  for different values of $C_0$ and panel (b) the conformations of the vesicle for $C_0=0.3$.}
\label{graph41}
\end{figure}

  {\it{Aggregation due to isotropic curvature generating proteins:}}  The spontaneous  curvature generated by proteins is accounted via Eqn.~\ref{Helfrich1} by assigning $C_0>0$ to the vertices occupied by the proteins. The clustering kinetics  and the corresponding conformations when $\phi_\%=30$, $p_\%=20$  are shown in Fig.~\ref{graph41}. As can be seen in  Fig.~\ref{graph41}(a), at early times (MC steps $\le 10^5$) the growth curve is nearly independent of the value of $C_0$. For MC steps $\ge 10^5$  the  protein cluster size saturates at a  value that decreases with increasing $C_0$.   The membrane deformations due to curvature active proteins are observed only when these proteins form a cluster.  When the cluster size reaches a threshold, at a value which depend on  $C_0$, membrane starts to deform  and this is reflected in the diffusion and growth of the clusters. In Sec~\ref{results21}  we have seen  bud  formation,  induced by  line tension,  takes place only when $\phi_\% \le p_\%$. Here we show that such buds can be  formed even when $\phi_\% > p_\%$ if the proteins are curvature active.   Earlier such  effects,  of spontaneous curvature on  vesicle conformations,  have been studied only  with direct  lipid-lipid interaction induced clustering~\cite{sunil:2001}. 
    
  {\it{Aggregation due to anisotropic curvature generating proteins:}} In order to study the role of anisotropic curvature induction in protein clustering we introduce curvature effects through the anisotropic curvature Hamiltonian  given in Eqn.~\ref{hanis}.  We concentrate on the conformations of the membrane when $\phi_\%=30$, $p_\%=20$ and $\kappa_{\bot}=0$.  It was shown earlier that, at these concentrations of anisotropic proteins  with $\epsilon_{LL}=0$   protein fields cannot induce large scale aggregation through  membrane curvature mediated interaction alone~\cite{Ramakrishnan:2013}  and  $\mathbf{H}_{p\phi}$ interactions are necessary for them to aggregate.  

\begin{figure}[h!]
\centering
\includegraphics[scale=0.5]{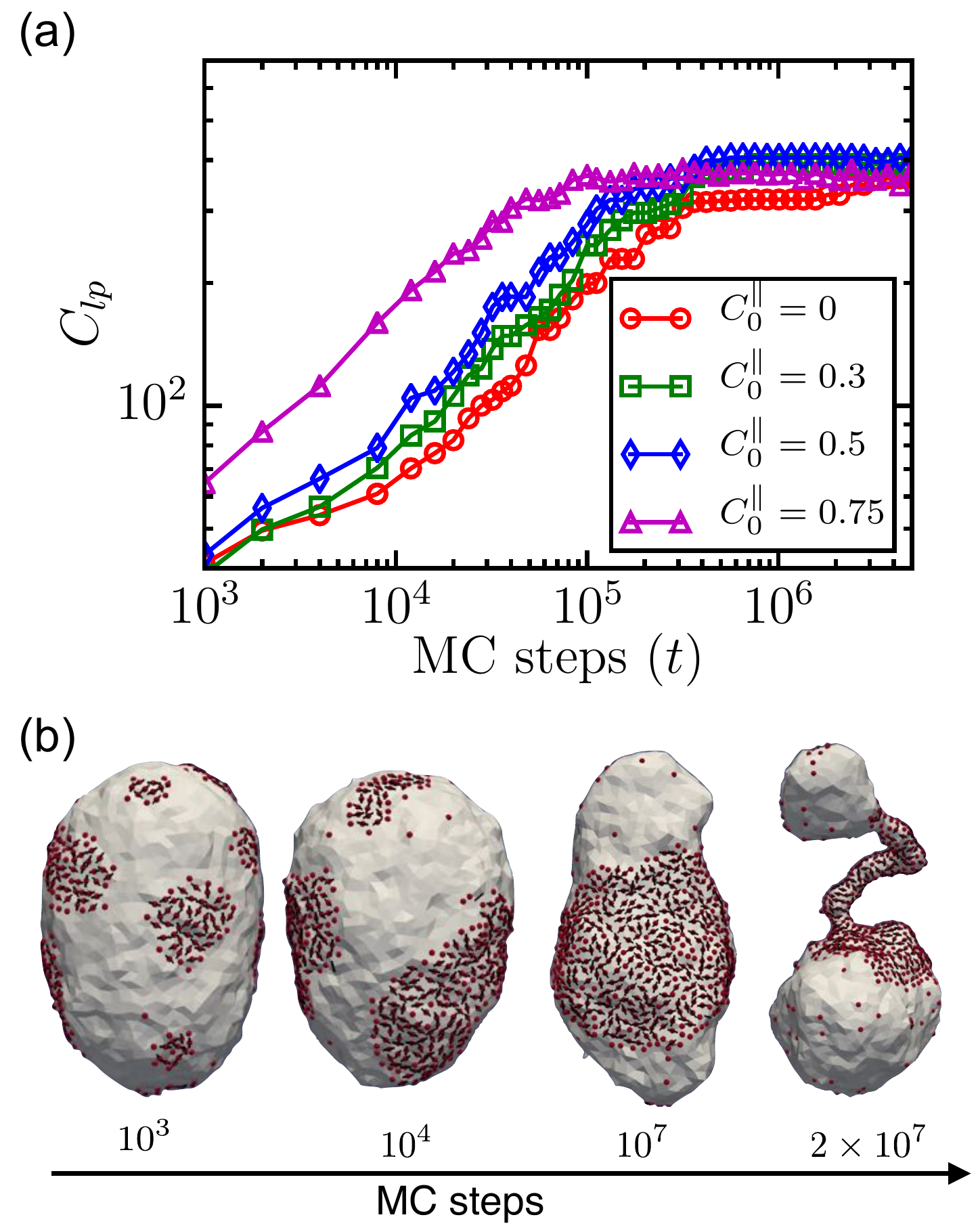}
\caption{Cluster formation in presence of  anisotropic  curvature generating proteins as a function of time when  $J_\phi=0$, $J_{p\phi} =2$,  $\epsilon_{LL}=0$  and $\kappa_{\parallel}=5$.  (a) $C_{lp}$ for different values of  $C_0^{\parallel}$.  (b)  The vesicle conformations  for induced curvature  $C_0^{\parallel}=0.75$.}
\label{graph4}
\end{figure}

The cluster formation for different values of $C_0^{\parallel}$ are shown in Fig.~\ref{graph4}(a).  The largest protein cluster size $C_{lp}$ at short time scales are found to be strongly dependent on the induced curvature. For example,   $C_{lp}$   for $C_0^{\parallel}=0.75$,  $\epsilon_{LL}=0$, exhibits the fastest growth,  showing  that  the proteins can enhance the aggregation through a  membrane mediated interaction between the protein fields. The membrane conformations  in the presence of curvature active proteins  when $J_{p\phi}=2$ are shown in Fig.~\ref{graph4}(b). Though no  explicit orientational interaction between the proteins are included, the proteins in a domain tend to align due to  similar induced curvature.   At early time the  clusters grow faster  than those seen for $C_0^{\parallel}=0$ (refer Fig.~\ref{conf1}(a)). At later times, these domains coalesce resulting in larger clusters until they are big enough to deform the membrane to form tubular structures. Such a large scale aggregation is not observed in the case of proteins that induce isotropic curvature as the domain induced budding occurs at an earlier stage. 

\section{Conclusions}

   We  propose a Monte Carlo based multicomponent vesicle model that  includes the effect of lipid-lipid, lipid-protein and protein-protein interactions to study (1) the kinetics of protein induced membrane domain formation and (2) the domain induced conformational changes of the membrane.  Our  study finds that  the lipid-protein interaction induced aggregations to  be significantly different from that resulting from  lipid-lipid interactions, with  the former showing faster cluster formation. The cluster sizes resulting from lipid-protein  interactions depend on the protein concentration and on their interaction strength. For small protein fractions ($p_\% < 20$) and low lipid-protein  interaction strengths ($J_{p\phi} < 2$), the absence of line tension at the domain boundaries lead to fast clustering kinetics.  At low  co-lipid  protein  composition ratio  and high  lipid-protein  interactions  the effect of line tension is prominent and   budding of membrane domains  and slowing down of domain growth is observed. 
 
 Our simulations  explored the effect of lipid-protein interactions on protein clustering and  showed that even at small protein concentrations it is possible to get  complete protein clustering. We also examined the role of the structural properties of the proteins on the protein aggregation. An explicit protein-protein interaction which prefers parallel alignment of proteins is shown to reduces diffusion of clusters and limits budding of domains. The  clustering of curvature active proteins is shown to have  significant dependence on the  anisotropy of the induced curvature, with stronger anisotropy suppressing  budding and  favoring larger cluster formation.

\appendix

\section{Scaling of domains}\label{appendix1}

Scaling assumption implies that the distance between  domains $d$  should scale the same way as the size of the domain size  itself.    For circular domains of radius $R$,  this would imply that   $d \propto R$.  If the domain coalescence takes place through diffusion of domains  then the coalescence time $t_c$  scales as  $d^2\propto Dt_c$, where $D$ is the diffusion coefficient.   Rouse dynamics implies that  $D\propto 1/R^2$, i.e., $D$ is inversely proportional to the  number of vertices in the domain.    Thus $d^2 \propto t_c/R^2$  which leads to  $R^4\propto t_c$.  Therefore   $R \propto t^{1/4}$ and the size of the domains scales as  $t^{1/2} $.

\nocite{*}
%

\end{document}